\documentclass[12pt]{article}
\usepackage{amsmath}
\usepackage{amscd}
\usepackage{amssymb}
\usepackage{array}

\begin{document}
\rightline{CERN-TH/2000-260}
\newcommand{\Z}{\mathbb{Z}}
\newcommand{\R}{\mathbb{R}}
\newcommand{\C}{\mathbb{C}}
\newcommand{\g}{\mathcal{G}}
\newcommand{\A}{\mathcal{A}}
\newcommand{\I}{\mathcal{I}}
\newcommand{\Ca}{\mathcal{C}}
\newcommand{\N}{\mathbb{N}}
\newcommand{\La}{\mathcal{L}}
\newcommand{\Ha}{\mathbb{H}}

\vskip 1cm

  \centerline{\LARGE \bf SPINOR ALGEBRAS}
\vskip 2cm

\centerline{ R. D'Auria$^\dagger$, S. Ferrara$^\bullet$,  M. A.
Lled\'o$^\dagger$ and V. S. Varadarajan$^\star$.}

\vskip 1cm

\centerline{\it $^\dagger$ Dipartimento di Fisica, Politecnico di
Torino,} \centerline{\it Corso Duca degli Abruzzi 24, I-10129
Torino, Italy, and} \centerline{\it INFN, Sezione di Torino,
Italy.}
 \medskip

\centerline{\it $^\bullet$ CERN, Theory Division, CH 1211 Geneva
23, Switzerland, and } \centerline{Laboratori Nazionali di
Frascati, INFN, Italy.}

\medskip

\centerline{\it $^\star$ Department of Mathematics, University of
California, Los Angeles,} \centerline{\it Los Angeles, CA
90095-1555, USA.}

\vskip 1cm

\begin{abstract}

We consider  supersymmetry algebras in space-times with arbitrary
signature and minimal number of spinor generators. The
interrelation between super Poincar\'e and super conformal
algebras is elucidated. Minimal super conformal algebras are seen
to have as bosonic part a classical semimisimple algebra naturally
associated to the spin group. This algebra, the Spin$(s,t)$
algebra, depends  both on the dimension and on the signature of
space time. We also consider maximal super conformal algebras,
which are classified by the  orthosymplectic algebras.


\end{abstract}

\vfill\eject

\section{Introduction}

In recent times the extension of Poincar\'e and conformal
superalgebras to orthosymplectic algebras has been considered with
a variety of purposes. In particular the role of osp($1|32, \R$)
and osp($1|64, \R$) as minimal superalgebras containing the
conformal algebras in 10 and 11 dimensions  (or the anti de Sitter
algebra in 11 and 12 dimensions) has been considered in view of
possible generalizations of M theory \cite{ hw, df, vv, to, to2,
gu2,
 fp, ho, bv, we} and of string theory to F-theory \cite{va}.
The contractions of orthosymplectic algebras are used in the study of BPS
branes \cite{ agit, to, fp2,  cgmv, du}.

In the present paper  we address the more general question of
whether such extensions are possible for space-times with Lorentz
group SO($s,t$). Space-times with more than one time direction
have been studied in order to unify duality symmetries of string
and M theories \cite{ba, hu} and
 to explore  BPS states in two-times physics \cite{ ba, adg}.
A theory based on the gauging of orthosymplectic algebras has been
suggested as a non perturbative definition of M-theory \cite{ho}.

Supersymmetric extensions of Poincar\'e and conformal
 (or anti de Sitter) algebras in higher dimensional spaces  have been considered
in the literature \cite{na, st, to, vv, vp, bv}. Our analysis
embraces all possible dimensions and signatures, so we will make
contact with the previous investigations.

We first consider $N=1$  super Poincar\'e algebras for arbitrary
space time signature and dimension,  extending the usual
classification of supersymmetries in any dimension \cite{na}. We
then compute the orthosymplectic superalgebras containing
so($s,t$) as a subalgebra of the symplectic algebra. The embedding
we look for is such that the symplectic fundamental representation
is an irreducible spinor representation when restricted to the
orthogonal algebra. Orthosymplectic superalgebras are seen to
contain Poincar\'e supersymmetry, either as a subalgebra or as a
Wigner-Inon\"u contraction. This generalizes the fact that the
M-theory superalgebra can be seen, either as a contraction of
osp($1|32, \R$) or as a subalgebra of osp($1|64, \R$).

The paper is organized as follows. In Section 2 we review properties of
 spinors and Clifford algebras for arbitrary signature and dimension
and set up the notation for the rest of the paper. We also provide
the symmetry properties of the morphisms which allow us the
classification of space-time superalgebras. In Sections 3 and 4
Poincar\'e and conformal supersymmetry are studied in a uniform
way. In Section 5 the orthosymplectic algebras and their
contractions to centrally extended super Poincar\'e and super
translation  algebras are studied.  In Section 6
 we introduce the concept of orthogonal symplectic and linear spinors which,
together with the reality properties allows us to associate a real
simple algebra from the classical series to the Spin group (called
 Spin($s,t$)-algebra). In Section 7 we show that the minimal
super conformal algebras are supersymmetric extensions of the
Spin($V$)-algebra. A maximal superalgebra with the same number of
odd generators is always an orthosymplectic algebra. In Section 8
we summarize our results and retrieve the examples of Minkowskian
signature.

\section{Properties of spinors of SO($\mathbf{V}$)\label{seclif}}

Let $V$ be a real vector space of dimension $D=s+t$ and
$\{v_\mu\}$ a basis of it. On $V$ there is a non degenerate
symmetric bilinear form which in the basis is given by the matrix
 $$\eta_{\mu\nu}={\rm diag}(+,\dots (s \;{\rm times})\dots,
+,-,\dots (t \;{\rm times})\dots, -).$$

We consider the group Spin($V$), the unique double covering of the
connected component of ${\rm SO}(s,t)$ and its spinor
representations. A spinor representation of Spin$(V)^\C$ is an
irreducible complex representation whose highest weights are the
fundamental weights corresponding to the right  extreme nodes in
the Dynkin diagram. These do not descend to representations of
SO$(V)$. A spinor type representation is any irreducible
representation that doesn't descend to SO$(V)$. A spinor
representation of Spin$(V)$ over the reals  is an irreducible
representation over the reals whose complexification is a direct
sum of spin representations.

 Two parameters, the signature $\rho$
mod(8) and the dimension $D$ mod(8) classify the properties of the
 spinor representation. Through this paper we will use the
following notation, $$ \rho=s-t=\rho_0 +8n,\qquad D=s+t=D_0 +8p,$$
where $\rho_0, D_0= 0,\dots 7$. We set $m=p-n,$ so
\begin{eqnarray*}s&=&\frac{1}{2}(D +\rho)=\frac{1}{2}(D_0+\rho_0)+8n+4m,\\
t&=&\frac{1}{2}(D-\rho)=\frac{1}{2}(D_0-\rho_0)+4m.\end{eqnarray*}

The signature  $\rho$ mod(8) determines if the spinor
representations are real ($\R$), quaternionic  ($\Ha$) or complex
($\C$) type.

The dimension $D$ mod(8) determines the nature of the
Spin($V$)-morphisms of the spinor representation $S$. Let $g\in
{\rm Spin}(V)$ and let  $\Sigma(g):S\longrightarrow S$  and
$L(g):V\longrightarrow V$ the spinor and vector representations of
$l\in{\rm Spin}(V)$ respectively. Then a
 map $A$
$$ A: S\otimes S
\longrightarrow \Lambda^k ,$$
 where $\Lambda^k=\Lambda^k(V)$ are
the $k$-forms on $V$, is a Spin($V$)-morphism if
$$A(\Sigma(g)s_1\otimes \Sigma(g)s_2)=L^k(g)A(s_1\otimes s_2). $$

\medskip

In the next subsections we analyze the properties of spinors for
arbitrary $\rho$ and $D$.

\subsection{Spinors and Clifford algebras\label{conju}}

 We
denote by $\Ca(s,t)$ the  Clifford algebra associated to $V$ and
$\eta$. It is defined as the real associative algebra generated by
the symbols $\I, \Gamma_\mu $ with relations \begin{equation}
\Gamma_\mu\Gamma_\nu+\Gamma_\nu\Gamma_\mu=  2\eta_{\mu\nu}\I,
\label{clial}\end{equation}
 and with $\I$ the unit element.

Let  $\C(p)$ be  the algebra of $p\times p$ complex matrices. The
complexification of the Clifford algebra, $\Ca(s,t)^\C\simeq
\Ca(t,s)^\C$, is
 isomorphic  to $\C(2^{D/2})$  for $D$ even  and to  $\C(2^{(D-1)/2})\oplus \C(2^{(D-1)/2})$ for $D$ odd.
  The real Clifford algebras are isomorphic to certain matrix
  algebras. They are classified by $\rho=s-t$ mod(8) (see
\cite{ch, bou, co, tr, vn, cb}). Notice that $D$ and $\rho$ have always the same
parity. We list the results in Table \ref{clifford}, where we have
used the following notation: $\mathbf{2}\times E=E\oplus E$,
  $\R(p)$ and $\C(p)$ mean the algebra of $p\times p$ matrices
 with entries in the real or complex numbers respectively. $\Ha(p)$
 instead means the set of $p\times p$ complex matrices satisfying the
 quaternionic condition

 \begin{equation}
M^*=-\Omega M\Omega\label{quat}
\end{equation}
where $\Omega$ is the symplectic metric. This means that $p$ is
 even and that $M$ can be written as a $p/2\times p/2$ matrix
 whose entries are quaternionic. Using the two dimensional complex representation
 of the quaternions we recover the previous description.
We stress that all the algebras appearing in Table \ref{clifford}
are taken as real algebras. The real dimension of the Clifford
algebra is in all cases $2^D$.

\begin{table}[ht]
\begin{center}
\begin{tabular} {|c ||c| c|c|c|}
\hline $\rho$ even &0 & 2 &4& 6\\ \hline $\Ca(s,t)$& $\R(2^{D/2})
$ &$\R(2^{D/2}) $ &$\Ha(2^{D/2-1}) $ &$\Ha(2^{D/2-1} )$ \\
\hline\hline $\rho$ odd &1 & 3 &5& 7\\ \hline $\Ca(s,t)$&
$\mathbf{2}\times\R(2^{(D-1)/2})$ &$\C(2^{(D-1)/2})$
&$\mathbf{2}\times\Ha(2^{(D-1)/2})$ &$\C(2^{(D-1)/2}) $ \\ \hline
\end{tabular}
\caption{Clifford algebras}\label{clifford}
\end{center}
\end{table}

We consider a representation of the Clifford algebra in a vector
space $S$ of dimension $2^{D/2}$ for $D$ even and $2^{(D-1)/2}$
for $D$ odd, as  given by Table \ref{clifford}. This
representation is faithful except for $\rho=1,5$ mod(8). We will
denote by $\gamma_\mu$ the images of the generators $\Gamma_\mu$
by this representation.  From Table \ref{clifford} one can see
also when these matrices are  real, quaternionic or just complex.
$S$ is then a real, quaternionic or complex vector space.

It is clear that in general $\Ca(s,t)$ and $\Ca(t,s)$ are not
isomorphic. However, the Clifford algebras have a natural $\Z_2$
grading, being the degree of $\Gamma_\mu$ equal to one. The
relations (\ref{clial}) are homogeneous in this degree. The even
(degree zero) part $\Ca^+(s,t)$ is a subalgebra generated by
products of an even number of elements of the basis $\Gamma_\mu$.
It is then true
 that   $\Ca^+(s,t)\simeq  \Ca^+(t,s)$. The Lorentz generators are
 products of two elements, so it follows trivially  that
 so$(s,t)\simeq $ so$(t,s)$. This will be important since we are in
 fact interested in the irreducible representations of Spin($V$).

For $D$ odd the  representation $S$ of the Clifford algebra
is irreducible under Spin($V$). It is a spinor representation.
 For $D$ even, it splits into two
irreducible spinor  representations (called Weyl or chiral spinors)
$S=S^+\oplus S^-$ of half the dimension.

We consider first the odd cases. Since for our purposes only
$|\rho_0|$ is important, we will have  up to two possible Clifford
algebras in each case.

\paragraph {$|\rho_0|=1$.} The Clifford algebras are the ones of
$\rho_0=1,7$. We see that $\rho=1$ gives directly a real
representation of real dimension $2^{(D-1)/2}$.

\paragraph {$|\rho_0|=3$.} The two possibilities are  $\rho_0=3,5$. $\rho_0=5$
gives a quaternionic representation of complex dimension
$2^{(D-1)/2}$.

 \paragraph {$|\rho_0|=5$.} As the case $|\rho_0|=3$.

\paragraph {$|\rho_0|=7$.} As the case $|\rho_0|=1$.

\bigskip
\bigskip

We consider now the even cases.

 \paragraph {$|\rho_0|=0$.} There is only  one possibility,  $\rho_0=0$.
 The representation is real of dimension $2^{D/2}$. The projections
 on $S^\pm$ are also real. This is because the projectors are
$$P^{\pm}=\frac{1}{2}(1\pm \gamma_{D+1})$$ where  $\gamma_{D+1}=
\gamma_1\cdots \gamma_D$, which is also real.

\paragraph {$|\rho_0|=2$.} The two possibilities are  $\rho_0=2,6$.
$\rho=2$ has a real representation, and $\rho=6$ has a
quaternionic representation. But the projectors in each  case are
not real nor quaternionic, $$P^{\pm}=\frac{1}{2}(1\pm
i\gamma_{D+1})$$ so the representations $S^\pm$ are just complex.

\paragraph {$|\rho_0|=4$.} There is only  one possibility,  $\rho_0=4$.
 The representation is quaternionic of complex dimension $2^{D/2}$. The projectors are
$$P^{\pm}=\frac{1}{2}(1\pm \gamma_{D+1}),$$ which is quaternionic,
so $S^\pm$ are also quaternionic representations.

  \paragraph {$|\rho_0|=6$.} As the case  $|\rho|=2$.

\bigskip
\bigskip

In Table \ref{realprop} we summarize  all these properties
together with the real dimension of the spinor representation.

\begin{table}[ht]
\begin{center}
\begin{tabular} {|c |c| c||c|c|c|}
\hline $\rho_0$(odd) &real dim($S$) & reality &$\rho_0$(even)
&real dim($S^{\pm}$) & reality\\ \hline \hline 1& $2^{(D-1)/2}$
&$\R$ & 0 &$2^{D/2-1}$&$\R$ \\ \hline 3& $2^{(D+1)/2}$ &$\Ha$& 2
&$2^{D/2}$&$\C$ \\ \hline 5& $2^{(D+1)/2}$ &$\Ha$& 4
&$2^{D/2}$&$\Ha$\\ \hline 7& $2^{(D-1)/2}$ &$\R$ & 6
&$2^{D/2}$&$\C$ \\ \hline
\end{tabular}
\caption{Properties of spinors}\label{realprop}
\end{center}
\end{table}

\bigskip

Space-time supersymmetry algebras are real superalgebras. The odd
generators are in spinor representations of the Lorentz group, so
we need to use real spinor representations. For  each case, real
quaternionic or complex, we use  an irreducible real spinor
representation, with the dimension indicated in Table
\ref{realprop}.

\paragraph{Real case, $\mathbf{\rho_0=0,1,7}$.}

Let $S$ be a finite dimensional complex vector space. A
conjugation $\sigma$ is a $\C$-antilinear map $\sigma:S\rightarrow
S$, $$\sigma(as_1 +bs_2)= a^*\sigma(s_1) +b^*\sigma(s_2), \qquad
a,b\in \C,\; s_i\in S,$$ such that $\sigma^2=\I$. Let  $S$ be the
vector space of an irreducible spinor representation of
${\rm Spin}(V)$. In this case there is a conjugation $\sigma$ that
 commutes with ${\rm Spin}(V)$, $$\sigma(gs)=g\sigma(s),
\quad g\in {\rm Spin}(V),$$ and then ${\rm Spin}(V)$ acts on the
real vector space $S^\sigma=\{s\in S| \sigma(s)=s\}$. The spinor
representation is an irreducible  representation of type $\R$.

\paragraph{Quaternionic case, $\mathbf{\rho_0=3,4,5}$.}
A pseudoconjugation is an antilinear map on $S$ such that $\sigma^2=-\I$.  $S$
 has necessarily even dimension.
If we have a real Lie algebra with an irreducible representation,
one can prove
 that it is of quaternionic type if and only if
 there exists a pseudoconjugation commuting with the action of the Lie
 algebra. So a quaternionic representation of ${\rm Spin}(V)$ has a
 pseudoconjugation $\sigma$. The condition
$\sigma(gs)=g\sigma(s)$ is equivalent, in a certain basis of
$S\simeq \C^{2n}$, to (\ref{quat}).

 Let $S$ be a quaternionic representation of ${\rm Spin}(V)$. We
take $\tilde S\simeq S\oplus S\simeq S\otimes W$, with $W=\C^2$.
On $S\otimes W$ we can define a conjugation
$\tilde\sigma=\sigma\otimes\sigma_0$, with $\sigma_0$ a
pseudoconjugation on $W$. In a basis of $W$   we can  always
choose $\sigma_0(w)=\Omega w^*$, with $\Omega$
$$\Omega=\begin{pmatrix} 0 &1\\-1&0\end{pmatrix}.$$ The biggest
group that commutes with $\sigma_0$ is $\rm{SU}(2)\simeq
\rm{USp}(2)\simeq \rm{SU}^*(2)$ , so we have that ${\rm
Spin}(V)\otimes {\rm SU}(2)$ commutes with $\tilde\sigma$ and has
a real representation on $${\tilde S}^\sigma=\{t\in \tilde S|
\tilde\sigma(t)=t\}.$$ We note at this point that there is a
smaller group, $ \rm{SO}^*(2)\simeq \rm{SO}(2)\simeq U(1)$
contained in $\rm{SU}(2)$. It will play a role in the construction
of superalgebras.

\paragraph{Complex case, $\mathbf{\rho_0=2,6}$.} The representation
 of the Clifford algebra
$\Ca(s,t)$ on $S=S^+\oplus S^-$ for $\rho_0=2$ is real. This means
that it has a conjugation which commutes with the action of
$\Ca(s,t)$. For $\rho_0=6$ the Clifford algebra is quaternionic,
which means that it has a pseudoconjugation. Nevertheless,  the
orthogonal group ${\rm Spin}(s,t)$ is isomorphic to
${\rm Spin}(t,s)$, so we can use the Clifford algebra $\Ca(t,s)$
which has $\rho_0=2$ and a conjugation.

We conclude then that for $\rho_0=2,6$ there is a conjugation
$\sigma$  on $S$ commuting with the action of ${\rm Spin}(V)$. If
follows that there is a representation of ${\rm Spin}(V)$ on the
real vector space  $S^\sigma$.

In particular, we have that $\sigma(S^\pm)=S^\mp$. We can define
an action of U(1) on $S$, $$e^{i\alpha}(s^+\oplus
s^-)=e^{i\alpha}s^+\oplus e^{-i\alpha}s^-.$$ This action commutes
also with $\sigma$, so it is defined on $S^\sigma$.

\bigskip

The groups SU(2) and U(1) appearing in the quaternionic and
complex case respectively  are referred to as R-symmetry groups.

\subsection{Spin$\mathbf{(V)}$-morphisms \label{raja}}

The symmetry properties of the Spin$(V)$-morphisms $$ S\otimes S
\longrightarrow \Lambda^k $$ depend on $D$ mod(8),  and are listed
in Table \ref{morphisms}. We put -1 if the morphism is
antisymmetric , +1 if it is symmetric and leave it  blank if no
symmetry properties can be defined. Notice that one can restrict
$k$ to $2k+1\leq D$ if $D$ is odd and to $2k\leq D$ if $D$ is even
since $\Lambda^k\simeq \Lambda^{(D-k)}$ are isomorphic as
Spin($V$)-modules. This table can be obtained exactly as table
1.5.1 in \cite{de}, using the formalism of \cite{de}.
\begin{table}[ht]
\begin{center}
\begin{tabular} {|c |c|c||c|c|}
\hline \multicolumn{1}{|c|}{$D$} &\multicolumn{2}{| c||}{$k $
even}&\multicolumn{2}{|c|}{$k $ odd}\\\hline\hline & morphism &
symmetry &morphism&symmetry\\
  \hline
  0& $S^\pm\otimes S^\pm\rightarrow\Lambda^k$&$(-1)^{k(k-1)/2}$& $S^\pm\otimes
  S^\mp\rightarrow\Lambda^k$&\\\hline
  1& $S\otimes S\rightarrow\Lambda^k$&$(-1)^{k(k-1)/2}$& $S\otimes
  S\rightarrow\Lambda^k$&$(-1)^{k(k-1)/2}$\\\hline
  2& $S^\pm\otimes S^\mp\rightarrow\Lambda^k$&& $S^\pm\otimes
  S^\pm\rightarrow\Lambda^k$&$(-1)^{k(k-1)/2}$\\\hline
  3& $S\otimes S\rightarrow\Lambda^k$&$-(-1)^{k(k-1)/2}$& $S\otimes
  S\rightarrow\Lambda^k$&$(-1)^{k(k-1)/2}$\\\hline
  4& $S^\pm\otimes S^\pm\rightarrow\Lambda^k$&$-(-1)^{k(k-1)/2}$& $S^\pm\otimes
  S^\mp\rightarrow\Lambda^k$&\\\hline
  5& $S\otimes S\rightarrow\Lambda^k$&$-(-1)^{k(k-1)/2}$& $S\otimes
  S\rightarrow\Lambda^k$&$-(-1)^{k(k-1)/2}$\\\hline
  6& $S^\pm\otimes S^\mp\rightarrow\Lambda^k$&& $S^\pm\otimes
  S^\pm\rightarrow\Lambda^k$&$-(-1)^{k(k-1)/2}$\\\hline
  7& $S\otimes S\rightarrow\Lambda^k$&$(-1)^{k(k-1)/2}$& $S\otimes
  S\rightarrow\Lambda^k$&$-(-1)^{k(k-1)/2}$\\\hline

\end{tabular}
\caption{Properties of morphisms.}\label{morphisms}
\end{center}
\end{table}

 Let $S^\vee$ be the dual space of $S$ and let
 $C_\pm:S\rightarrow S^\vee$ be the map intertwining two equivalent  representations of the Clifford algebra,
namely
\begin{alignat*}{2}
C_+^{-1}\gamma_\mu C_+ &=&\gamma_\mu^T, \qquad &{\rm for} \; D=1\;
{\rm mod}(4)\\ C_-^{-1}\gamma_\mu C_- &=&-\gamma_\mu^T, \qquad
&{\rm for} \; D=3\; {\rm mod}(4)\\ C_\pm^{-1}\gamma_\mu C_\pm
&=&\pm\gamma_\mu^T, \qquad &{\rm for} \; D\;{\rm even}.
\end{alignat*}
Notice that $C_\pm$ defines a map $S\otimes S\rightarrow \C$. This
map has the property of being a Spin($V$)-morphism, so its
symmetry properties can be deduced from Table \ref{morphisms}. In
terms of a basis of $S$, $\{e_\alpha\}$, and its dual,
$\{e^\vee_\alpha\}$, both the morphism and the intertwining map
are expressed as a matrix ${C_\pm}_{\alpha \beta}$ called the
charge conjugation matrix \cite{tr,vn,vp}.

In the even case, $S=S^+\oplus S^-$. For $D=0,4$ the morphisms  $S\otimes S\rightarrow \C$ are
block diagonal ( $S^\pm\otimes S^\pm\rightarrow \C$), so the charge conjugation matrices must be both
 symmetric or both antisymmetric.  For $D=2,6$ the morphisms are  off diagonal, ($S^\pm\otimes S^\mp\rightarrow \C$),
 so the charge conjugation matrices can have simultaneously different symmetry properties.  In fact, we have
\begin{eqnarray*}
D&=&0\; {\rm mod}(8)\qquad C_\pm^T=C_\pm\\
D&=&2\; {\rm mod}(8)\qquad C_\pm^T=\pm C_\pm\\
D&=&4 \;{\rm mod}(8)\qquad C_\pm^T=-C_\pm\\
D&=&6 \;{\rm mod}(8)\qquad C_\pm^T=\mp C_\pm.
\end{eqnarray*}
For $D$ odd we have
\begin{eqnarray*}
D&=&1 \;{\rm mod}(8)\qquad C_+^T=C_+\\
D&=&3 \;{\rm mod}(8)\qquad C_-^T=- C_-\\
D&=&5 \;{\rm mod}(8)\qquad C_+^T=-C_+\\
D&=&7 \;{\rm mod}(8)\qquad C_-^T=C_-.
\end{eqnarray*}

For arbitrary $k$ we have that the gamma matrices
$$\gamma^{[\mu_1,\dots \mu_k]}=\frac{1}{k!}\sum_{s\in S^k}{\rm sig}(s)\gamma^{\mu_{s(1)}}\cdots
 \gamma^{\mu_{s(k)}}$$
are a map $S\rightarrow \Lambda^k\otimes S$. Composing it with
$\I\otimes C$ we obtain a map $S\rightarrow \Lambda^k\otimes
S^\vee$, which defines a map $S\otimes S\rightarrow \Lambda^k$.
This map is a Spin($V$)-morphism, and in the same basis as before
is given by $$\gamma^{[\mu_1,\dots \mu_k]}_{\alpha\beta}=
\frac{1}{k!}\sum_{s\in
S^k}{\rm sig}(s){\gamma^{\mu_{s(1)}}}_\alpha^{\beta_1}
{\gamma^{\mu_{s(2)}}}_{\beta_1}^{\beta_2}\cdots
 {\gamma^{\mu_{s(k)}}}_{\beta_{k-1}}^{\beta_k}C_{\beta_k \beta}.$$

\paragraph {A note on Majorana spinors.}

Consider the orthogonal group $\rm {SO}(s,t)$. For $\rho_0 = 1,7$
the spinors in  the representation $S^\sigma$, of dimension
$2^{(D-1)/2}$,
 are  called Majorana spinors.
For $\rho_0=0$ the  spinors in $(S^\pm)^\sigma$ (of dimension
$2^{D/2-1}$)
  are called Majorana-Weyl. For
$\rho_0 = 2, 6$ the space of Majorana spinors is  $(S^+\oplus
S^-)^\sigma$, of dimension $2^{D/2}$.

For $\rho_0 = 3,5$ the quaternionic spinors in $S$ are called
pseudoMajorana spinors.  For $\rho_0 =
 4$, the Weyl spinors are themselves quaternionic and they are
 called pseudoMajorana-Weyl spinors.

The space of Majorana spinors is a real vector space  and the
space of  pseudoMajorana spinors is a  quaternionic vector space
\cite{fv, vp, tr, vn}.

\section{Poincar\'e supersymmetry}

The Poincar\'e group of a space $V$ of signature $(s,t)$ is the
group ${\rm ISO}(s,t))= {\rm SO}(s,t)\circledS {\rm T}^{s+t} $. We
consider super Poincar\'e algebras with  non extended
supersymmetry ($N=1$). The anticommutator of the odd generators
(spinor charges) is in the representation Sym($S\otimes S$).
One can decompose it into  irreducible representations under the group
${\rm Spin}(V)$. It is a fact that only antisymmetric tensor
representations will appear. Poincar\'e supersymmetry requires the
presence of the vector representation in this decomposition to
accommodate the momenta $P_\mu$. Another way of expressing this is
by saying that there must be a morphism $$S\otimes
S\longrightarrow V$$ which is symmetric. This can be read from
Table \ref{morphisms}. In the table, complex representations are
considered.
 Since the Poincar\'e superalgebra is a real superalgebra,
care should be exercised when interpreting
it in the different cases of real, quaternionic and complex
spinors. We will deal separately with these cases.

\bigskip

\paragraph{Real case.}
 The most general form of the anticommutator of two spinor generators is
\begin{equation}\{Q_\alpha,Q_\beta\}=\sum\limits_{k} \gamma^{[\mu_1\cdots
\mu_k]}_{(\alpha \beta)}Z_{[\mu_1\cdots \mu_k]},
\label{sutrans}\end{equation} where $Z_{[\mu_1\cdots\mu_k]}$ are even
generators.
 In the sum there appear only the terms that are
symmetric with respect to $\alpha$ and $\beta$; we indicate it by
$(\alpha \beta)$.

  If the term $\gamma^\mu_{(\alpha\beta)}$ appears, then a super
Poincar\'e algebra exists. The rest of the $Z$  generators can be
taken to commute among themselves and with the odd generators and
transform appropriately with the Lorentz generators. We have then
the maximal ``central extension"\footnote{ Except for $k=0$, the
generators $Z$ are not central elements, since they do not commute
with the elements of the Lorentz group. They are central only in
the super translation algebra. It is nevertheless customary to
call them ``central charges".} of the super Poincar\'e algebra.

For $\rho_0=0$, since the Weyl spinors are real one can have a
chiral superalgebra. The vector representation should appear then
in the symmetric product ${\rm Sym}(S^\pm\otimes S^\pm)$. This
happens only for $D_0=2$ ($\rho$ and $D$ have the same parity). If
we consider non chiral superalgebras, where both $S^\pm$ are
present, also the values $D_0=0, 4$ are allowed.

For $\rho_0=1,7$,
 we have $D_0= 1,3$.

\paragraph{Quaternionic case.}
 The most general anticommutator of two
spinor charges is
\begin{equation}\{Q^i_\alpha,Q^j_\beta\}=\sum\limits_{k} \gamma^{[\mu_1\cdots
\mu_k]}_{[\alpha \beta]}Z^0_{[\mu_1\cdots \mu_k]}\Omega^{ij}+
\sum\limits_{k} \gamma^{[\mu_1\cdots \mu_k]}_{(\alpha
\beta)}Z^I_{[\mu_1\cdots \mu_k]}\sigma_I^{ij}.
\label{sutrans2}\end{equation} $\sigma_I^{ij}$ are the (symmetric)
Pauli matrices, $i,j=1,2$, $I=1,2,3$ (We have multiplied them by
the invariant antisymmetric metric $\Omega^{ij}$). If we demand
that the momentum $P_\mu$ is  a singlet under the full R-symmetry
group $\rm{SU}(2)\simeq \rm{Usp}(2)$, then the
$\gamma_{\alpha\beta}^\mu$ must be antisymmetric and the momentum
appears  in the first term (singlet) of the r.h.s. of
(\ref{sutrans2}).

For $\rho_0=3,5$, this happens if  $D_0=5,7$. The only even case
is $\rho_0=4$. A  chiral  superalgebra exists for $D_0=6$

If we restrict the R-symmetry group to $\rm{SO}^*(2)$, there is
also an invariant symmetric metric, $\delta^{ij}$. In the
anticommutator (\ref{sutrans2}) we can consider terms  like
$$\sum\limits_{k} \gamma^{[\mu_1\cdots \mu_k]}_{(\alpha
\beta)}Z_{[\mu_1\cdots \mu_k]}\delta^{ij}.$$ The
$\gamma_{\alpha\beta}^\mu$ must be symmetric to appear in such
term. For $\rho_0=3,5$, this happens if $D_0=1,3$. For $\rho_0=4$
and $D_0=2$ a chiral superalgebra exists.

 For $\rho_0=4$ and $D_0=0,4$ one can have non chiral
superalgebras.

\paragraph{Complex case.} This is the case for $\rho_0=2,6$. The spinor
charges are in the representation $S^+\oplus S^-$ and we will denote them by
$(Q_\alpha,Q_{\dot\alpha})$. In the anticommutator there are three pieces,
$$\{Q_\alpha,Q_\beta\}, \quad \{Q_{\dot\alpha},Q_{\dot\beta}\},
\quad \{Q_{\alpha},Q_{\dot\alpha}\}, $$
and it is clear that only the last one is invariant under the R-symmetry group
U(1). Then there must be a morphism
$$S^+\otimes S^-\longrightarrow \Lambda^1.$$
  This happens in the
cases $D_0=0,4$.

\bigskip
\bigskip
We summarize these results in Table \ref{iso}.

\begin{table}[ht]
\begin{center}
\begin{tabular} {|c|l||c|l|}
\hline
 $(D_0, \rho_0)$  &ISO($s,t$)& $(D_0,\rho_0)$ &ISO($s,t$)\\\hline\hline
(2,0) & ISO($1+8n+4m,1+4m$) & (1,1) & ISO($1+8n+4m,4m$)\\\hline

(0,2) &  ISO($1+8n+4m,-1+4m$) & (1,3)&ISO($2+8n+4m,-1+4m$)\\\hline

(4,2)&ISO($3+8n+4m,1+4m$)& (3,3) & ISO($3+8n+4m,4m$)\\\hline

(2,4) & ISO($3+8n+4m,-1+4m$)& (1,5) & ISO($3+8n+4m,-1+4m$)\\\hline

 (6,4) & ISO($5+8n+4m,1+4m$)& (3,5) & ISO($4+8n+4m,-1+4m$)\\\hline

 (0,6) & ISO($3+8n+4m,-3+4m$)  &(3,1) & ISO($2+8n+4m,1+4m$)\\\hline

 (4,6) & ISO($9+8n+4m,3+4m$)  &(5,3) & ISO($4+8n+4m,1+4m$) \\\hline

(0,0)$^\dagger$& ISO($8n+4m,4m$) &(7,3) &
ISO($5+8n+4m,2+4m$)\\\hline (0,4)$^\dagger$&
ISO($2+8n+4m,-2+4m$)&(5,5) & ISO($5+8n+4m,4m$) \\\hline
(4,0)$^\dagger$& ISO($2+8n+4m,2+4m$)  &(7,5) & ISO($6+8n+4m,1+4m$)
\\\hline
 (4,4)$^\dagger$&  ISO($4+8n+4m,4m$)& (1,7) & ISO($4+8n+4m,-3+4m$) \\\hline
 & & (3,7)& ISO($5+8n+4m,-2+4m$) \\\hline

\end{tabular}
\caption{Poincar\'e groups with supersymmetric extensions}\label{iso}
\end{center}
\end{table}
The values of $m,n$ are such that $s,t\geq 0$. We mark with
``$\,\dagger\,$'' the non chiral superalgebras. We note that for
standard space-time signature, ${\rm ISO}(D-1,1)={\rm ISO}(\rho+1 +8n,1)$,
all super Poincar\'e algebras are present for $D=0,\dots,7$
mod(8).

\section{Conformal supersymmetry\label{consu}}

The conformal group of a vector space  $V$  of signature
$(s-1,t-1)$ is the group  of coordinate transformations that leave
the metric invariant up to a  scale change. This group is
isomorphic to SO$(s,t)$, a simple group, for $D\geq 3$. The
Poincar\'e group ISO$(s,t)$ is a subgroup
 of the conformal group. In a space with the standard Minkowski signature
$(s-1,1)$,
 the conformal group  is the simple
group SO$(s,2)$. It is also the anti de Sitter group in dimension
$s+1$.

 A simple superalgebra $\A=\A_0\oplus
 \A_1$ satisfies necessarily
 \begin{equation}
 \{\A_1, \A_1\}=\A_0
 \label{property}
 \end{equation}
 We look for  minimal simple superalgebras (with minimal number of even generators) containing
 space-time conformal symmetry in its even part. The odd generators are in a
 spinor representation $S$ of ${\rm Spin}(s,t)$, and
 all the even generators should appear in the
  right hand side of the anticommutator of the spinor charges, which
 is in the Sym($S\otimes S$) representation. As we did in the case of
 Poincar\'e supersymmetry, we decompose it with respect to  ${\rm Spin}(s,t)$.
The orthogonal generators are in the   antisymmetric
 2-fold representation, so we
 should  look for morphisms
$$S\otimes S\longrightarrow \Lambda^2$$ with the appropriate
symmetry properties for each signature and dimension. The
discussion is as for  Poincar\'e supersymmetry, but with $k=2$ in
Table \ref{morphisms}.

For the real case the matrices should be symmetric. We have
$\rho_0=0$ with $D_0=4$ and $\rho_0=1,7$ with $D_0=3,5$. For the
quaternionic case the matrices should be antisymmetric if we
demand that the orthogonal generators appear as a  singlet under
the SU(2) R-symmetry.
 We have $\rho_0=4$ with $D_0=8$ and $\rho_0=3,5$ with $D_0=1,7$.
If the R-symmetry is restricted to $\rm{SO}^*(2)$ the singlet is
$\delta^{ij}$, while the $\rm{SO}^*(2)$ generator is
$\Omega^{ij}$.  Then we have $\rho_0=4$ with $D_0=4$ and
$\rho_0=3,5$ with $D_0=3,5$

For the complex case, if we demand that the orthogonal generators
are singlets under U(1), the matrices should be in  $S^+\otimes
S^-$, which is invariant under the U(1) R-symmetry group. We have
$\rho_0= 2,6$ with $D_0=2,6$. For  $\rho_0= 0$ and $D_0=2,6$ we
have a superalgebra containing the orthogonal group in its even
part provided we take two spinors, one in $S^+$ and the other in
$S^-$. $\rho_0= 4$ and $D_0=2,6$ is a similar case, but the
spinors in $S^\pm$ should have also an SU(2) index.  We may also
consider the cases $\rho_0=2, 6$ and $D_0=4$ where the U(1)
invariance is not present. Then, the orthogonal generators are in
the anticommutator ${\rm Sym} (S^+\otimes S^+)$.

When the morphism is such that the orthogonal generators are in
the r.h.s. of the anticommutator of the odd generators,
 the biggest simple group that one can consider is the one
generated by all the symmetric matrices. This is the symplectic group ${\rm Sp}(2n,\R)$ where $2n$
is the real dimension of the spinor charge. As we will see there is a
superalgebra with bosonic part ${\rm sp}(2n,\R)$, one of the
 orthosymplectic algebras. In the quaternionic
case, we observe that if the morphism to $\Lambda^2$ is
antisymmetric (symmetric), then the morphism to $\Lambda^0=\C$ is
symmetric (antisymmetric). It follows from (\ref{sutrans2}) that
in this case the orthogonal group times SU(2) is a subgroup of the
symplectic group. In the complex case the orthogonal group will
come multiplied by U(1) (unless $D_0=4$). In these cases, SU(2)
and U(1) respectively are groups of automorphisms of the
supersymmetry algebra.

  The results are
summarized in Table \ref{con} with the same conventions as in
Table \ref{iso}. We mark with  ``$\,\dagger\,$'' the cases that lead to
non chiral superalgebras.

\begin{table}
\begin{center}
\begin{tabular} {|c|l|l|}
\hline
 $(D_0,\rho_0)$  &SO($s,t$)& Sp($2n,\R))$\\\hline\hline
 (0,4)& SO($2+8n+4m,-2+4m)\times$SU(2) &Sp($2^{4(n+m)})$\\\hline
 (2,0)$^\dagger$ & SO($1+8n+4m,1 +4m$) &Sp($2^{1+4(n+m)} $)\\\hline
 (2,2)&SO($2+8n+4m,4m)\times $U(1) &Sp($2^{1+4(n+m)})$\\\hline
 (2,4)$^\dagger$ & SO($3+8n+4m,-1+4m)\times$SU(2) &Sp($2^{2+4(n+m)}$)\\\hline
 (2,6) & SO($4+8n+4m,-2+4m)\times$U(1)&Sp($2^{1+4(n+m)}$)\\\hline
 (4,0) & SO($2+8n+4m,2+4m$) & Sp($2^{1+4(n+m)})$ \\\hline
  (4,2) & SO($3+8n+4m,1+4m$)&Sp($2^{2+4(n+m)})$\\\hline
  (4,4) & SO($4+8n+4m,4m)\times $SO$^*(2)$&Sp($2^{2+4(n+m)})$\\\hline
  (4,6) &SO($5+8n+4m,-1+4m$)&Sp($2^{2+4(n+m)})$\\\hline
  (6,0)$^\dagger$ & SO($3+8n+4m,3+4m$) &Sp($2^{1+4(n+m)}$)\\\hline
  (6,2) & SO($4+8n+4m,2+4m)\times$U(1) &Sp($2^{3+4(n+m)})$\\\hline
   (6,4)$^\dagger$ & SO($5+8n+4m,1+4m)\times$SU(2) &Sp($2^{2+4(n+m)}$)\\\hline
   (6,6) & SO($6+8n+4m,4m)\times $U(1)&Sp($2^{3+4(n+m)})$\\\hline\hline

  (1,3) & SO($2+8n+4m,-1+4m)\times $SU(2)&Sp($2^{1+4(n+m)})$\\\hline
  (1,5) & SO($3+8n+4m,-2+4m)\times$SU(2)&Sp($2^{1+4(n+m)})$\\\hline
    (3,1) & SO($2+8n+4m,1+4m$)&Sp($2^{1+4(n+m)})$\\\hline
(3,3) & SO($3+8n+4m,4m)\times $SO$^*(2)$
&Sp($2^{2+4(n+m)})$\\\hline (3,5) & SO($4+8n+4m,-2+4m)\times
$SO$^*(2)$ &Sp($2^{2+4(n+m)})$\\\hline (3,7) & SO($5+8n+4m,-2+4m$)
&Sp($2^{1+4(n+m)})$\\\hline
    (5,1) &SO($3+8n+4m,2+4m$)&Sp($2^{2+4(n+m)})$\\\hline
 (5,3) & SO($4+8n+4m,2+4m)\times $SO$^*(2)$
&Sp($2^{3+4(n+m)})$\\\hline (5,5) & SO($5+8n+4m,4m)\times
$SO$^*(2)$ &Sp($2^{3+4(n+m)})$\\\hline (5,7) & SO($6+8n+4m,-1+4m$)
&Sp($2^{2+4(n+m)})$\\\hline
    (7,3) & SO($5+8n+4m,2+4m)\times$SU(2)&Sp($2^{4+4(n+m)})$\\\hline
   (7,5) &SO($6+8n+4m,1+4m)\times$SU(2)&Sp($2^{4+4(n+m)})$\\\hline
\end{tabular}
\caption{Orthogonal groups and their symplectic
 embeddings.}\label{con}
\end{center}
\end{table}

The case of SO(2,2) would naively correspond to an embedding in
${\rm Sp}(2,\R)$. This is obviously not true and the reason is
that O(2,2) is not simple, so property (\ref{property}) doesn't
hold. In fact, since ${\rm SO}(2,2)\simeq {\rm SO}(2,1)\times{\rm
SO}(2,1)$,
 we have that
$\rm {Sp}(2,\R)\simeq{\rm SO}(2,1)$, one of the simple factors.

\section{The orthosymplectic algebra  and space time supersymmetry}

We recall here the definition of the  orthosymplectic superalgebra
osp($N|2p,\R$) \cite{nrs, bg, gu2}. Consider the  $\Z_2$-graded
vector space $E=V\oplus H$, with dim$( V)=N$ and dim$(H)=2p$.
${\rm End}(E)$ is a super Lie algebra in the usual way, with the
even part ${\rm End}(E)_0={\rm End}(V)\oplus{\rm End}(H)$. In
terms of an homogeneous basis, an element of End$(E)$ is a real
matrix $$\begin{pmatrix} A_{N\times N}&B_{2p\times N}\\ C_{N\times
2p}&D_{2p\times 2p} \end{pmatrix}.$$ Then $${\rm
End}(E)_0=\begin{pmatrix} A&0\\0&D
\end{pmatrix},\qquad {\rm End}(E)_1=\begin{pmatrix} 0&B\\C&0
\end{pmatrix},$$ with the usual bracket
\begin{equation}
[a,b]=ab-(-1)^{g(a)g(b)}ba,\qquad a,b\in {\rm End}(E).\label{cr}
\end{equation}
($g=0,1$ will denote the grading on both spaces, $E$ and
End$(E)$).

Consider on $E$ a non degenerate bilinear form $F$  such that
$F(u,v)=(-1)^{p(u)p(v)}F(v,u)$ and $F(u,v)=0$ for $u\in E_0, v\in
E_1$. Then, there exists an homogeneous basis where
$$F=\begin{pmatrix}\Omega_{N\times N}& 0\\0 &\Omega_{2p\times
2p}\end{pmatrix},$$ with
\begin{eqnarray*}&\Omega_{2p\times 2p}^2=-\I, \qquad \Omega_{2p\times
2p}^T=-\Omega_{2p\times 2p}\\
&\Omega_{N\times N}^T=\Omega_{N\times N}.
\end{eqnarray*}

The orthosymplectic algebra osp($N|2p,\R$) is the set of real
 $(N+2p)\times(N+2p)$ matrices $a$ satisfying
 $$ a^TF+Fa=0,$$
 with bracket (\ref{cr}). The even part is ${\rm so}(N)\oplus{\rm sp}(2p)$, and the generators
 of the odd part are in the fundamental representation $(N,2p)$. It is a simple superalgebra, so in
 particular,
\begin{equation}
\{{\rm osp}(N|2p,\R)_1,{\rm osp}(N|2p,\R)_1\}={\rm
so}(N)\oplus{\rm sp}(2p,\R). \label{simple}\end{equation}

Given the results of Section \ref{consu}, the orthosymplectic
super algebras are the supersymmetric extensions of the conformal
group  of space-time. We take $N=1$, and $2p$ (the dimension of
the symplectic group) according to Table \ref{con}. The defining
representation of the symplectic group is the corresponding spinor
representation of the orthogonal subgroup.

\bigskip

The symplectic algebra sp$(2p)$ has a maximal subalgebra
sl($p,\R)\oplus$so(1,1). The fundamental representation of
sp($2p$) decomposes as $$
\begin{CD}\mathbf{2p}@>>{\rm sl}(p,\R)\oplus{\rm so}(1,1)>(\mathbf{p},\frac{1}{2})\oplus
 (\mathbf{p}',-\frac{1}{2})\end{CD},$$
where $\mathbf{p}'$ is the dual representation to $\mathbf{p}$.
The decomposition of the adjoint representation
 is
 \begin{eqnarray*}\begin{CD}{\rm Sym}(\mathbf{2p}\otimes\mathbf{2p})@>>{\rm sl}(p,\R)
\oplus{\rm so}(1,1)> ({\rm Sym}(\mathbf{p}\otimes\mathbf{p}),1)
\oplus ({\rm adj}_{\rm sl(p)},0)\oplus\end{CD}\\
\oplus(1,0)\oplus({\rm Sym}(\mathbf{p}'\otimes\mathbf{p}'),-1)
\end{eqnarray*} This defines an so(1,1), Lie algebra grading of $\La={\rm
sp}(2p)$
 \begin{equation}\La_{{\rm sp}}=\La_{{\rm sp}}^{+1}\oplus \La_{{\rm sp}}^0\oplus
   \La_{{\rm sp}}^{-1},\label{3grading}\end{equation}
 where the superindices are the so(1,1) charges. The direct sums here are
 understood as vector space sums, not as Lie algebra sums. By the properties
 of the grading, $\La_{{\rm sp}}^{+1}$ and
 $\La_{{\rm sp}}^{-1}$ are abelian subalgebras.

 The orthogonal group SO($s,t$) contains as a subgroup
 ISO($s-1,t-1$). In the algebra, the adjoint of so$(s,t)$ contains a singlet under
 SO$(s-1,t-1)$. The corresponding so(1,1)-grading is like
 (\ref{3grading}), and in fact they coincide when the orthogonal
 algebra is seen as a subalgebra of the symplectic one. For the orthogonal
 case we have
$$\La_{{\rm o}}^{+1}=\{P_\mu\},\qquad
\La_{{\rm o}}^0={\rm so}(s-1,t-1)\oplus {\rm o}(1,1), \qquad
\La_{{\rm o}}^{-1}=\{K_\mu\} $$ where $P_\mu$ and $K_\mu$ are
so($s-1,t-1$) vectors satisfying $$[P_\mu, P_\nu]=[K_\mu,
K_\nu]=0.$$ $\La_{{\rm o}}^0$ contains the orthogonal generators
$M_{\mu\nu}\in \rm {so}(s-1,t-1)$ and the dilatation $D$. $P_\mu$
can be identified with the momenta of ISO($s-1,t-1$), and $K_\mu$
are the conformal boost generators.

 When we consider the supersymmetric extension of Sp($2p$) as the
 orthosymplectic algebra osp$(1|2p,\R)$, the previous grading is
 extended and we have the decomposition
$$
 \La_{{\rm osp}}=\La_{{\rm osp}}^{+1}\oplus\La_{{\rm osp}}^{+1/2}\oplus\La_{{\rm osp}}^0\oplus
 \La_{{\rm osp}}^{-1/2}\oplus\La_{{\rm osp}}^{-1}
 $$
 where $\La_{{\rm osp}}^{\pm 1/2}$
  contain the odd generators of the
 superalgebra, which are in the fundamental representation of
 sp($2p$). This representation decomposes as
$$ \begin{CD}\mathbf{2p}@>>{\rm sl}(p,\R)\oplus{\rm so}(1,1)>
(\mathbf{p},+\frac{1}{2})\oplus
(\mathbf{p}',-\frac{1}{2})\end{CD}$$ so

$$\La_{{\rm osp}}^{+ 1/2}=\{Q_\alpha\},\qquad \La_{{\rm osp}}^{-
1/2}=\{S_\alpha\}.$$

 It is
 important to remark that since the signature $\rho$ is the same for so$(s,t)$
 and so$(s-1,t-1$), the spinors have the same reality properties.
 Furthermore, the irreducible spinor of ${\rm so}(s,t)\subset{\rm sp}(2p)$ decomposes
 into
 two irreducible spinors of ${\rm so}(s-1,t-1)\subset{\rm sl}(p)$ with opposite
 grading. These are usually denoted as the $Q$ and $S$ spinors of
 the super conformal algebra.

 When $D=s+t$ is even, the irreducible spinor $S^\pm$ of
 ${\rm so}(s,t)$ decomposes into two spinors of
 ${\rm so}(s-1,t-1)$ of opposite chiralities,
 $$S^+_D\longrightarrow Q^+_{D-2} \oplus S^-_{D-2}.$$
 (the superindex here indicates chirality, not the so(1,1)-grading).
 Since one has the morphism \cite{de}
 $$Q^\pm_{D-2}\otimes V\longrightarrow S^\mp_{D-2},$$
 then  the commutator of a
 charge of a certain chirality with $K_\mu$ or $P_\mu$ must give a
 charge with the opposite chirality.

  The subalgebra
 $\La^{+1}\oplus\La^{+1/2}$ is a nilpotent subalgebra of
 osp$(1|2p,\R)$, which in fact is the maximal central extension
 of the super translation algebra. The full set of central charges
 transforms, therefore, in the symmetric representation of
 sl($p,\R$) while the odd charges transform in the fundamental
 representation of the same group.  We observe that the
 orthosymplectic algebra has twice the number of odd generators
 than the super Poincar\'e algebra.

 \subsection{Contractions of the orthosymplectic algebra}

 The Poincar\'e algebra can also be obtained from an orthogonal
 algebra by an Inon\"u-Wigner contraction
 $$\begin{CD}{\rm so}(s,t+1)@>>{\rm contraction}>{\rm iso}(s,t),\end{CD}$$
  as a generalization of the well known case of the  anti
  De-Sitter gorup in $D-1$ dimension
 $$\begin{CD}{\rm SO}(D-2,2)@>>{\rm contraction}>{\rm ISO}(D-2,1)\end{CD}.$$
 In fact, the same Poincar\'e algebra can be obtained also by the
 contraction
$$\begin{CD}{\rm so}(s+1,t)@>>{\rm contraction}>{\rm iso}(s,t).\end{CD}$$

 The contraction is defined as follows. Let $T_{[AB]}$ be the
 generators of ${\rm so}(s,t+1$) or ${\rm so}(s+1,t$) , $A,B=1,\dots D',\;\; D'=s+t+1$. Let
 $\mu,\nu=1,\dots D$ and consider the decomposition
 $T_{[\mu\nu]}, T_{[\mu,D']}$. We define
 $T'_\mu=\frac{1}{e}T_{[\mu,D']}$ and take the limit $e\rightarrow
 0$ in the algebra while keeping finite the  generators
 $T_{[\mu\nu]}, T_{[\mu,D']}$. The result is the algebra of the Poincar\'e
 group  ${\rm ISO}(s,t)$ with $P_\mu=T_{[\mu,D']}$.

 We consider now the following contraction of the orthosymplectic
 superalgebra. The generators of the bosonic subalgebra
  $Z_{[\mu_1\cdots \mu_k]}$ appear in the r.h.s. of (\ref{qq})

\begin{equation}\{Q_\alpha,Q_\beta\}=\sum\limits_{k} \gamma^{[\mu_1\cdots
\mu_k]}_{\alpha \beta}Z_{[\mu_1\cdots \mu_k]}, \qquad
\mu_i=1,\dots D .\label{qq}\end{equation} (Only the $\gamma$'s
with the appropriate symmetry will appear). We set
\begin{eqnarray*} Z_{[\mu_1\cdots \mu_k]}&\longrightarrow&
\frac{1}{e}Z_{[\mu_1\cdots \mu_k]}\\
Q&\longrightarrow&
 \frac{1}{\sqrt{e}}Q.
 \end{eqnarray*}
 We obtain a superalgebra with bosonic part totally abelian.

Consider a symplectic group containing an orthogonal group in
dimension $D$ and signature $\rho$  according to Table \ref{con},
and the contraction of the orthosymplectic algebra as explained
above. We can decompose the odd and even generators with respect
to the orthogonal subgroup $(D-1, \rho+1)$
 or $(D-1, \rho-1)$. Interpreting $Z_{[\mu D]}$ as the momentum in dimension $D-1$
  ($\mu$ taken only from 1 to $D-1$),
  the algebra is then seen to be
 the maximal central extension of the  super translation algebra  in $(D-1, \rho\pm 1)$.

 If all the symplectic generators are contracted except the
 generators of the orthogonal group,
 $$Z_{[\mu\nu]}\mapsto Z_{[\mu\nu]}, \qquad \mu,\nu=1,\dots D-1,$$
  then one obtains a super Poincar\'e algebra. It has ``central
 extension'', but it is not maximal since the generators of the
 orthogonal group SO($s,t)$, $Z_{[\mu\nu]}$, are not commuting and
 do not appear in the right hand side of (\ref{qq}).

The
 spinor representations of the orthogonal
 group in ($D,\rho$)  behave differently when decomposing with respect to the
  orthogonal  subgroup in ($D-1, \rho\pm 1$),
 depending on $\rho$. For the complex spinors we have that for $D\rightarrow D-1$
 \begin{eqnarray*}S_D^\pm&\mapsto &S_{D-1},\qquad \qquad \quad \;\; D\;\; {\rm even}\\
 S_D&\mapsto &S_{D-1}^+\oplus S_{D-1}^-,\qquad D\;\; {\rm odd}.
 \end{eqnarray*}
 Over the reals,  the representation may or may not remain
 irreducible. We make the analysis first for ($D-1, \rho +1$).
 The representation remains
 irreducible for $\rho_0=0,1,2,4$ while for $\rho_0= 3,5,6,7$ it
 splits into two spinor representations, so the super Poincar\'e
 algebra obtained has $N=2$ supersymmetry. More precisely, for
 $\rho_0=3,7$ we get two spinors of different chirality, so we have (1,1)
 supersymmetry, while for $\rho=5,6$ we get $N=2$ supersymmetry.

 The orthogonal group for  ($D-1, \rho -1$) is isomorphic to the orthogonal group for
 ($D-1, -\rho +1$), so the decomposition of the representations under
 $\rho\mapsto \rho-1$ can be
formulated as
a decomposition of the type
 $\rho'\mapsto \rho'+1$ with $\rho'=-\rho$. It is then enough to write the decompositions
$\rho\mapsto \rho+1$. We give them in
  Table \ref{sred}.

\begin{table}[ht]
\begin{center}
\setlength{\extrarowheight}{4pt}
\begin{tabular} {|c|c|c|}
\hline
  $\rho_0$ &  $(\rho,D)\rightarrow (\rho+1,D-1)$ &
  Reducibility
  \\\hline\hline
 0 & $S_\R^+\rightarrow S_\R$&
irreducible\\\hline
  1& $S_\R\rightarrow S_\C^+ \oplus \overline{S^+_\C}$& irreducible \\\hline
2& $S_\C^+\rightarrow S_\Ha$ & irreducible\\\hline 3 &
$S_\Ha\rightarrow S_\Ha^+ \oplus S_\Ha^-$& reducible\\\hline
4&$S^+_\Ha\rightarrow S_\Ha$& irreducible\\ \hline 5 &
$S_\Ha\rightarrow S_\C^+ \oplus\overline{ S_\C^-}$&
reducible\\\hline 6 & $S_\C^+\rightarrow S_\R\oplus S_\R$&
reducible\\\hline 7 & $S_\R\rightarrow S_\R^+ \oplus S_\R^-$&
reducible\\\hline
\end{tabular} \caption{Decomposition of spinors $(\rho,D)\rightarrow (\rho+1,D-1)$}\label{sred}
\end{center}
\end{table}

We can now apply these decompositions to the list given in Table
\ref{con}. The  super Poincar\'e algebra for $(D,\rho)$ could be
in principle obtained by contraction from two different
orthosymplectic algebras, the ones corresponding to orthogonal
groups $(D+1,\rho+1)$ or $(D+1,\rho-1)$. However, it may happen
that no one of them exists, as in  $(D_0,\rho_0)=(0,0)$ or
 that only one exists, as for
$(D_0,\rho_0)=(7,3),(7,5),(0,2),(2,2),(6,2),(6,6)$. The rest have
both possibilities.

  The orthosymplectic algebra corresponding to $(D,\rho)$
can be contracted in two different ways, as the bosonic orthogonal algebra.
 However these contractions do  not lead  necessarily
 to one of the Poincar\'e superalgebras listed in Table \ref{iso}, since we
imposed some restrictions on the algebras appearing in that table.

Let us note that in particular, the Poincar\'e algebras corresponding to the
physically interesting case, $\rho=D-2$, are all obtained by contraction. For $D=8,9$
the algebra obtained has extended ($N=2$) supersymmetry. For $D=6,10$ the algebras obtained
are non chiral.

\section{Orthogonal, symplectic and linear spinors}

We consider now morphisms  \cite{de,ac}$$S\otimes S\longrightarrow
\Lambda^0\simeq \C.$$ If a morphism of this kind exists, it is
unique up to a multiplicative factor. The vector space of the
spinor representation has then a bilinear form invariant under
Spin($V$).
 Looking at Table \ref{morphisms}, one can see that this morphism
 exists except for $D_0=2,6$, where instead a morphism
 $$S^{\pm}\otimes S^{\mp}\longrightarrow \C$$
 occurs.

 We shall call a spinor representation orthogonal if it has a
 symmetric, invariant bilinear form. This happens for $D_0=0,1,7$ and
 Spin$(V)^\C$ (complexification of Spin($V$)) is then  a subgroup
 of the complex orthogonal group ${\rm SO}(n,\C)$, where $n$ is the
 dimension of the spinor representation (Weyl spinors for $D$ even).
 The generators
 of ${\rm SO}(n,\C)$ are  $n\times n$ antisymmetric matrices. These are obtained
 in terms of the morphisms
$$S\otimes S\longrightarrow\Lambda^k,$$ which are  antisymmetric.
This gives  the decomposition of the adjoint representation of
${\rm SO}(n,\C)$ under the subgroup ${\rm Spin}(V)^\C$.
 In particular, for $k=2$ one obtains the generators of ${\rm Spin}(V)^\C$.

 A spinor representation is called symplectic if it has an
 antisymmetric, invariant bilinear form. This is the case for
 $D_0=3,4,5$. ${\rm Spin}(V)^\C$ is a subgroup of the symplectic group
 ${\rm Sp}(2p,\C)$, where $2p$ is the dimension of the
 spinor representation. The Lie algebra ${\rm sp}(2p,\C)$ is formed by all the
 symmetric matrices, so it is  given in terms of the
  morphisms $S\otimes S\rightarrow \Lambda^k$ which are symmetric.
 The generators
 of ${\rm Spin}(V)^{\C}$ correspond to $k=2$ and  are symmetric matrices.

 For $D_0=2,6$
 one has an
 invariant morphism
 $$B:S^{+}\otimes S^{-}\longrightarrow \C.$$
The representations $S^+$ and $S^-$ are
one the contragradient (or dual) of the other.
 The spin representations extend to  representations of the
 linear group ${\rm GL}(n,\C)$, which leaves the pairing  $B$ invariant. These
 spinors are called linear. Spin($V)^\C$ is a subgroup of the simple factor SL$(n,\C)$.

 These properties depend exclusively on the dimension. When combined with the
 reality properties, which depend on
 $\rho$, one obtains real groups embedded in ${\rm SO}(n,\C)$, ${\rm Sp}(2p,\C)$ and
 ${\rm GL}(n,\C)$ which have an action on the space of the real spinor
 representation $S^\sigma$. The real groups contain as a subgroup
 ${\rm Spin}(V)$.

We need first some general facts about real forms of simple Lie
algebras. Let $S$ be a complex vector space of dimension $n$ which
carries an irreducible representation of a complex Lie algebra
$\g$. Let $G$ be the complex Lie group associated to $\g$.  Let
$\sigma$ be a conjugation or a  pseudoconjugation on $S$ such that
$\sigma X\sigma^{-1}\in \g$ for all $X \in \g.$ Then  the map
$$X\mapsto X^\sigma=\sigma X\sigma^{-1}$$ is a conjugation of
$\g$. We shall write $$\g^\sigma=\{X\in \g|X^\sigma=X\}.$$
$\g^\sigma$ is a real form of $\g$. If $\tau=h\sigma h^{-1}$,
with $h\in \g$, $\g^\tau=h\g^\sigma h^{-1}$. $\g^\sigma=\g^{\sigma'}$
if and only if $\sigma'=\epsilon \sigma$ for $\epsilon$ a scalar with $|\epsilon|=1$;
in particular, if $\g^\sigma$ and $\g^\tau$ are conjugate by $G$,
$\sigma$ and $\tau$ are both conjugations or both pseudoconjugations. The conjugation can
also be defined on the group $G$, $g\mapsto \sigma g\sigma^{-1}$.

\subsection{Real forms of the classical Lie algebras}

We  describe  the real forms of the classical Lie algebras from this
point of view. (See also \cite{he}).

\paragraph{Linear algebra, sl($\mathbf{ S}$).}

\subparagraph{(a)} If $\sigma$ is a conjugation of $S$, then
there is an isomorphism  $S\rightarrow  \C^n$  such that $\sigma$
goes over to the standard conjugation of $\C^n$. Then
$\g^\sigma\simeq{\rm sl}(n,\R)$. (The conjugation acting  on gl$(n,\C)$ gives
 the real form gl($n,\R$)).

\subparagraph{(b)} If $\sigma$ is a pseudoconjugation and $\g$
doesn't leave invariant a non degenerate bilinear form, then there
is an isomorphism of $S$ with $\C^n$, $n=2p$ such that $\sigma$
goes over to $$(z_1,\dots,z_p,z_{p+1},\dots z_{2p})\mapsto
(z^*_{p+1},\dots z^*_{2p},-z^*_1,\dots,-z^*_p).$$ Then
$\g^\sigma\simeq {\rm su}^*(2p)$. (The pseudoconjugation acting in
on gl$(2p,\C)$ gives
 the real form ${\rm su}^*(2p)\oplus{\rm so}(1,1)$.)

To see this, it is enough to prove that $\g^\sigma$ does not leave
invariant any non degenerate hermitian form, so it cannot be of
the type su$(p,q)$. Suppose that $\langle\cdot ,\cdot\rangle$ is a
$\g^\sigma$-invariant, non degenerate hermitian form.  Define
$(s_1,s_2):=\langle\sigma (s_1),s_2\rangle$. Then $(\cdot ,\cdot)$
is bilinear and $\g^\sigma$-invariant, so it is also
$\g$-invariant.

\subparagraph{(c)} The remaining cases, following E. Cartan's
classification of real forms of simple Lie algebras, are
${\rm su}(p,q)$, where a non degenerate hermitian bilinear form is
left invariant. They do not correspond to a conjugation or
pseudoconjugation on $S$, the space of the fundamental
representation. (The real form of gl$(n,\C)$ is in this case u$(p,q)$).

\paragraph{Orthogonal algebra, so($\mathbf{S}$).} $\g$ leaves invariant a non degenerate,
symmetric bilinear form. We will denote it by $(\cdot,\cdot)$.

\subparagraph{(a)} If $\sigma$ is a conjugation preserving $\g$, one can prove that
 there is an isomorphism of
$S$ with $\C^n$ such that  $(\cdot,\cdot)$ goes to the standard
form and $\g^\sigma$ to ${\rm so}(p,q)$, $p+q=n$. Moreover, all
${\rm so}(p,q)$ are obtained in this form.

\subparagraph{(b)} If $\sigma$ is a pseudoconjugation preserving $\g$,
 $\g^\sigma$ cannot be
any of the ${\rm so}(p,q)$. By E. Cartan's classification, the only
other possibility is that $\g^\sigma\simeq {\rm so}^*(2p)$. There
is an isomorphism of $S$ with $\C^{2p}$ such that $\sigma$ goes to
$$(z_1,\dots z_p,z_{p+1},\dots z_{2p})\mapsto (z^*_{p+1},\dots
z^*_{2p},-z^*_{1},\dots -z^*_{p}).$$

\paragraph{Symplectic algebra, sp($\mathbf{S}$).} We denote by $(\cdot,\cdot)$ the symplectic
form on $S$. \subparagraph{(a)}If $\sigma$ is a conjugation
preserving $\g$, it is clear  that there is an isomorphism of $S$
with $\C^{2p}$, such that $\g^\sigma\simeq {\rm sp}(2p,\R)$.

\subparagraph{(b)}If $\sigma$ is a pseudoconjugation preserving
$\g$, then $\g^\sigma\simeq {\rm usp}(p,q)$, $p+q=n=2m, \;
p=2p',\; q=2q' $. All the real forms ${\rm usp}(p,q)$ arise in
this way. There is an isomorphism of $S$ with $\C^{2p}$ such that
$\sigma$ goes to $$(z_1,\dots z_m,z_{m+1},\dots z_{n})^t\mapsto
J_mK_{p',q'}(z^*_1,\dots z^*_m,z^*_{m+1},\dots z^*_{n})^t,$$ where
\begin{equation*}J_m=\begin{pmatrix}0&I_{m\times m}\\-I_{m\times
m}&0\end{pmatrix},\qquad K_{p',q'}=\begin{pmatrix}-I_{p'\times
p'}&0&0&0\\0&I_{q'\times q'}&0&0\\ 0&0&-I_{p'\times
p'}&0\\0&0&0&I_{q'\times q'}\end{pmatrix}.\end{equation*}

\bigskip

 At the end
of Section \ref{conju} we saw that there is  a conjugation on $S$
when  the spinors are real and a pseudoconjugation when they are
quaternionic (both denoted by $\sigma$). We have a group,
${\rm O}(n,\C)$, ${\rm Sp}(2p,\C)$ or
 ${\rm GL}(n,\C)$ acting on $S$ and  containing ${\rm Spin}(V)^\C$.  We note
that this group is minimal in the classical group series. If the
Lie algebra $\g$ of this group is stable under the conjugation
$$X\mapsto \sigma X\sigma^{-1}$$ then the real Lie algebra
$\g^\sigma$ acts on $S^\sigma$ and contains the Lie algebra of
${\rm Spin}(V)$. We shall call it the Spin($V$)-algebra.

Let $B$ be the space of ${\rm Spin}(V)^\C$-invariant bilinear forms
on $S$. Since the representation on $S$ is irreducible, this space
is at most one dimensional. Let it be one dimensional and let
 $\sigma$ be  a conjugation or a
pseudoconjugation and let $\psi\in B$. We define a conjugation
in the space $B$ as \begin{eqnarray*} B&\longrightarrow &B\\
\psi&\mapsto& \psi^\sigma\end{eqnarray*} $$
\psi^\sigma(v,u)=\psi(\sigma(v),\sigma(u))^*.$$ It is then
immediate that we can choose $\psi\in B$ such that
$\psi^\sigma=\psi$. Then if $X$ belongs to the Lie algebra
preserving $\psi$, so does $\sigma X\sigma^{-1}$.

\subsection{Spin($\mathbf{s,t}$)-algebras \label{realforms}}

We now  determine   the real Lie algebras  in each
case. All the possible cases must be studied separately. We start
with odd dimensions. All dimension and signature relations are
mod(8). In the following, a relation like ${\rm Spin}(V)\subseteq
G$ for a group $G$ will mean  that the image of ${\rm Spin}(V)$
under the spinor representation  is in the connected component of
$G$. The same applies for the  relation ${\rm Spin}(V)\simeq G$.

 \paragraph{Orthogonal spinors in odd dimension,  $\mathbf{D_0 = 1,7}$}

  \subparagraph{ Real spinors, $\mathbf{\rho_0= 1,7}$.} There is a
  conjugation $\sigma$ on $S$
   commuting with ${\rm Spin}(V)$. Then ${\rm Spin}(V)\subseteq {\rm SO}(S^\sigma)\simeq{\rm SO}(p,q)$. To
   determine $p$ and $q$, we look at the embedding of the maximal
   compact subgroup of ${\rm Spin}(V)$ into ${\rm SO}(p)\times {\rm SO}(q)$. We have three cases:

   \smallskip

   \noindent (a) If $\rho=D$ ($s$ or $t$ is zero), ${\rm Spin}(V)$ is compact
    and it is embedded in the compact orthogonal group,
$${\rm Spin}(V)\subseteq{\rm SO}(2^{(D-1)/2},\R),$$
     so  $p$ or $q$ is zero. This is clear
 since the lowest dimensional spinor type representation of
 ${\rm Spin}(V)$ is $\mathbf{2}^{(D-1)/2}$.

 \smallskip

 \noindent (b) If $s$ or $t$ is 1, then the maximal
 compact subgroup of ${\rm Spin}(V^D)$ is ${\rm Spin}(V^{D-1})$.
 Let $\varepsilon$ be the non trivial central element
 of ${\rm Spin}(V^D)$ which maps to the identity under the homomorphism
 ${\rm Spin}(V^D)\rightarrow {\rm SO}(V^D)$. Under
 the injection
 $${\rm Spin}(V^D)\longrightarrow {\rm SO}(S^\sigma)\simeq{\rm SO}(p,q),$$
 the central element $\varepsilon$ maps to $ -\I_{p+q}$. The
 compact subgroup of  ${\rm Spin}(V^D)$ maps into the maximal
 compact subgroup of ${\rm SO}(p,q)$, so that
 $${\rm Spin}(V^{D-1})\longrightarrow {\rm SO}(p)\times{\rm SO}(q).$$
 But the dimension of any spinor type  representation of ${\rm Spin}(V^{D-1})$ is
  bigger or equal than ${2}^{(D-1)/2-1}$. Since
 $\varepsilon$  maps to
 $-\I_{p}\oplus -\I_{q}$, both maps
 $${\rm Spin}(V^{D-1})\rightarrow {\rm SO}(p)\quad \rm{and}\quad  {\rm Spin}(V^{D-1})\rightarrow {\rm SO}(q)$$
 are spinor type  representations.  It follows that $p,q \geq {2}^{(D-1)/2-1}$, so $p=q=
 {2}^{(D-1)/2-1}$. So
$$ {\rm Spin}(V^D)\subseteq{\rm SO}(2^{(D-1)/2-1},{2}^{(D-1)/2-1}).$$

 \smallskip

\noindent (c) If $s,t\geq 2$, the maximal compact subgroup  of ${\rm Spin}(V^D)$ is
${\rm Spin}(s)\times {\rm Spin}(t)/(\varepsilon_s=\varepsilon_t)$,
where $\varepsilon_{s}$ and $\varepsilon_{t}$ are the central
elements in ${\rm Spin}(s)$ and $ {\rm Spin}(t)$ respectively, and
they must be identified with $\varepsilon$. The embedding of the
maximal compact subgroup must be $${\rm Spin}(s)\times
{\rm Spin}(t)/(\varepsilon_s=\varepsilon_t)\longrightarrow
{\rm SO}(p)\times {\rm SO}(q).$$

 The  spinor type  representation of ${\rm Spin}(s)\times {\rm Spin}(t)/(\varepsilon_s=\varepsilon_t)$
  of minimal dimension is
$\mathbf{2}^{(s-1)/2}\otimes\mathbf{2}^{t/2-1}$ if $s$ is odd and
$t$ even (only with a tensor product representation is possible to
identify $\varepsilon_s$ and $\varepsilon_t$). For the same reason
that in (b), we have that $p=q={2}^{(D-1)/2-1}$. So
$${\rm Spin}(V)\subseteq{\rm SO}(2^{(D-1)/2-1},{2}^{(D-1)/2-1}).$$ Low
dimensional examples are
 $${\rm Spin}(4,3)\subset {\rm SO}(4,4),\qquad {\rm Spin}(8,1)\subset{\rm SO}(8,8).$$

We  give now a more complicated example. Consider the group
${\rm Spin}(12,5)$. The spinor representation is {\bf 256}, and
should be embedded in the vector representation of ${\rm SO}(p,q)$,
$p+q=256$. We have the following decomposition
$$\begin{CD}\mathbf{256}@>>{\rm Spin}(12)\times{\rm Spin}(5)>\mathbf{(32^+,4)}\oplus
\mathbf{(32^-,4)}\end{CD}. $$ It follows that $p=q=128$, so the
group will be ${\rm SO}(128,128)$.

 Note that  the representations of
${\rm Spin}(12)$ and ${\rm Spin}(5)$ are quaternionic separately,
but when tensoring them a reality condition can be imposed.

Since there is no symmetric morphism $S\otimes S\rightarrow
\Lambda^2$ one cannot construct in this case
 a simple superalgebra containing
the orthogonal group.

\subparagraph{ Quaternionic spinors $\mathbf{\rho_0= 3,5}$.} We
have that ${\rm Spin}(V)$ commutes with a pseudoconjugation on $S$.
It then follows that
$${\rm Spin}(V)\subseteq{\rm SO}^*(2^{(D-1)/2}).$$ A low dimensional
example is $${\rm Spin}(6,1)\subset{\rm SO}^*(8),\qquad {\rm Spin}(5,2)
\subset{\rm SO}^*(8).$$

We explicitly compute another example, the group ${\rm Spin}(10,5)$
whose quaternionic spinor
 representation is $\bf 128$. We have the following decomposition
 $$\begin{CD}\mathbf{128}@>>{\rm Spin}(10)\times {\rm Spin}(5)>\mathbf{(16^+,4)}\oplus
 \mathbf{(16^-,4)}\end{CD}. $$

\paragraph{ Symplectic spinors in odd dimension, $\mathbf{D_0
=3,5}$.}

\subparagraph{Real spinors, $\mathbf{\rho_0= 1,7}$.}

 Since there is a  conjugation commuting with ${\rm Spin}(V)$,
$${\rm Spin}(V)\subseteq{\rm Sp}(2^{(D-1)/2},\R).$$ We have the low
dimensional examples $$ {\rm Spin}(2,1)\simeq {\rm SL}(2,\R), \qquad
\rm {Spin}(3,2)\simeq{\rm Sp}(4,\R).$$

\subparagraph{Quaternionic spinors, $\mathbf{\rho_0= 3,5}$.}

${\rm Spin}(V)$ commutes with a pseudoconjugation, so
${\rm Spin}(V)\subseteq {\rm USp}(p,q)$. We have three cases,

\smallskip

\noindent (a) If $s$ or $t$ is zero, then ${\rm Spin}(V)$ is
compact  and $${\rm Spin}(V)\subseteq {\rm USp}(2^{(D-1)/2}).$$ Low
dimensional examples are $${\rm Spin}(3)\simeq {\rm SU}(2),\qquad
{\rm Spin}(5)\simeq {\rm USp}(4). $$

\smallskip

\noindent (b) If $s$ or $t$ are 1, then the maximal compact
subgroup is ${\rm Spin}(V^{D-1})$ is embedded in ${\rm
USp}(p)\times{\rm USp}(q)$. The same reasoning as in the
orthogonal case can be applied here, and  $p=q=2^{(D-1)/2-1}$. So
$${\rm Spin}(V)\subseteq {\rm USp}(2^{(D-1)/2-1},2^{(D-1)/2-1}).$$
Low dimensional examples are $${\rm Spin}(4,1)\simeq{\rm
USp}(2,2).$$

\smallskip

\noindent (c) If $s,t\geq 2$, then $${\rm Spin}(s)\times
{\rm Spin}(t)/(\varepsilon_s=\varepsilon_t)\longrightarrow
{\rm USp}(p)\times {\rm USp}(q).$$ As before, $p=q={2}^{(D-1)/2-1}$
and so $${\rm Spin}(V)\subseteq
{\rm USp}(2^{(D-1)/2-1},2^{(D-1)/2-1}).$$

\bigskip

We analyze now the even dimensional cases.

\paragraph{Orthogonal spinors in even dimensions, $\mathbf{D_0 = 0}$.}

\subparagraph{Real spinors, $\mathbf{\rho_0=0}$.} The group $\rm
{Spin}(V)^\pm$ (projections of  ${\rm Spin}(V)$ with the chiral or
Weyl  representations)  commutes with a conjugation. Using the
same reasoning as in the odd case, we have that for a compact
group $${\rm Spin}(V)^\pm\subseteq {\rm SO}(2^{D/2-1}).$$ An example
of this is ${\rm Spin}(8)\simeq {\rm SO}(8)$\footnote {Notice that
for $D=8$ one has the phenomenon of triality.}.  For a non compact
group we have $${\rm Spin}(V)^\pm\subseteq
{\rm SO}(2^{D/2-2},2^{D/2-2}),$$ as for example
${\rm Spin}(4,4)\simeq {\rm SO}(4,4).$

 \subparagraph{Quaternionic spinors, $\mathbf{\rho_0=4}$.} We note
 that $s$ and $t$ are both even and that neither can be zero. ${\rm Spin}(V)^\pm$
 commutes with a pseudoconjugation, so $${\rm Spin}(V)^\pm\subseteq {\rm SO}^*(2^{D/2-1}).$$
  An example is ${\rm Spin}(6,2)\simeq {\rm SO}^*(8).$

\subparagraph{Complex spinors, $\mathbf{\rho_0=2,6}$.}  $s$ and
$t$ must be bigger than zero.  ${\rm Spin}(V)^\pm$ does not commute
with a conjugation or pseudoconjugation, since it is not real nor
quaternionic. It follows that there is no  real form of
${\rm SO}(2^{D/2-1},\C)$ containing ${\rm Spin}(V)^\pm$. We have
instead, using \cite{de} $${\rm Spin}(V)^\pm\subseteq
{\rm SO}(2^{D/2-1},\C)_\R,$$ which is also a simple real group.
(The suffix ``$\R$'' means that the complex group is considered as a
real Lie group). It cannot be seen as a real form of any complex
simple Lie group \cite{he}. As an example, $${\rm Spin}(7,1)\subset
{\rm SO}(8,\C)_\R.$$

\paragraph{Symplectic spinors  in even dimensions, $\mathbf{D_0 = 4}$.}

\subparagraph{Real spinors, $\mathbf{\rho_0=0}$.} $p$ and $q$ are
both even and neither can be zero. We have
$${\rm Spin}(V)^\pm\subseteq {\rm Sp}(2^{D/2-1},\R).$$ The lowest
dimensional case is not simple,
$${\rm Spin}(2,2)\simeq{\rm Sp}(2,\R)\times {\rm Sp}(2,\R).$$

\subparagraph{Quaternionic spinors, $\mathbf{\rho_0=4}$.}
${\rm Spin}(V)^\pm$ commute with a pseudoconjugation, so
${\rm Spin}(V)^\pm\subseteq {\rm USp}(p,q)$, $p+q=2^{D/2-1}$. Again,
the lowest dimensional case is semisimple, $$\rm
{Spin}(4)\simeq{\rm SU}(2)\times {\rm SU}(2).$$

 If $s$ or $t$ are zero we are in the compact case and
 $${\rm Spin}(V)^\pm\subseteq {\rm USp}(2^{D/2-1}).$$  The other possible
case is $s,t>0$, and then $s,t\geq 4$. As in the even case, we
have $p=q=2^{D/2-2}$, $${\rm Spin}(V)^\pm\subseteq
{\rm USp}(2^{D/2-2},2^{D/2-2}).$$

\subparagraph{Complex spinors, $\mathbf{\rho_0=2,6}$.} As in the
orthogonal case, no real form of ${\rm Sp}(2^{D/2-1},\C)$
containing ${\rm Spin}(V)^\pm$ exists. We have the embedding
$${\rm Spin}(V)^\pm\subseteq {\rm Sp}(2^{D/2-1},\C)_\R.$$ An example
is $${\rm Spin}(3,1)={\rm Sp}(2,\C)_\R\simeq {\rm SL}(2,\C)_\R.$$

\paragraph{Linear spinors, $\mathbf{D_0 = 2,6}$.}

\subparagraph{Real spinors, $\mathbf{\rho_0=0}$.}
${\rm Spin}(V)^\pm$ commutes with a conjugation, so one has an
embedding into the standard real form of the linear group,
$${\rm Spin}(V)^\pm\subset {\rm SL}(2^{D/2-1}, \R).$$ As an example, we
have ${\rm Spin}(3,3)^\pm\simeq {\rm SL}(4, \R).$

\subparagraph{Quaternionic spinors, $\mathbf{\rho_0=4}$.} The
representations $S^\pm$ are dual to each other and they commute
with a pseudoconjugation. They leave no bilinear form invariant.
If there were   an invariant hermitian form
$\langle\cdot,\cdot\rangle$ then one could define an invariant
bilinear form $$(s_1,s_2)=\langle\sigma s_1,s_2\rangle.$$ So the
only possibility is $${\rm Spin}(V)^\pm\subseteq
{\rm SU}^*(2^{D/2-1}).$$ A low dimension example is
${\rm Spin}(5,1)\simeq {\rm SU}^*(4)$.

\subparagraph{Complex spinors, $\mathbf{\rho_0=2,6}$.}  We will
denote by $(\cdot,\cdot)$ the ${\rm Spin}(V)$-invariant pairing
between $S^-$ and $S^+$. We remind that on $S=S^+\oplus S^-$ there
is a conjugation $\sigma$ commuting with the action of
${\rm Spin}(V)$ (see Section \ref{seclif}). It satisfies
$\sigma(S^\pm)=S^\mp$, so we can define a
${\rm Spin}(V)^\pm$-invariant sesquilinear form on $S^+$, $$\langle
s_1^+,s_2^+\rangle=(\sigma(s_1^+),s_2^+),\qquad s_i^+\in S^+.$$ By
irreducibility of the action of ${\rm Spin}(V)^\pm$, the space of
invariant sesquilinear forms is one dimensional. We can choose an
hermitian form as a basis, so it follows that
$${\rm Spin}(V)^\pm\subseteq {\rm SU}(p,q).$$ If $s$ or $t$ are zero
(compact case), then we have that $p,q\geq 2^{D/2-1}$, so either
$p$ or $q$ are zero and $${\rm Spin}(V)^\pm\subseteq
{\rm SU}(2^{D/2-1}).$$

If neither $p$ nor $q$ is zero, then $p,q\geq2$ and even. We have
that the embedding of the maximal compact subgroup must be
$${\rm Spin}(p)\times{\rm Spin}(q)/(\varepsilon_p=\varepsilon_q)
\subseteq {\rm S(U}(p)\times{\rm U}(q)).$$ So $p,q\geq
2^{p/2-1}\times 2^{q/2-1}=2^{D/2-2}$. It follows that
$${\rm Spin}(V)^\pm\subseteq {\rm SU}(2^{D/2-2},2^{D/2-2}).$$ We
have the low dimensional examples
$${\rm Spin}(6)\simeq{\rm SU}(4),\quad
{\rm Spin}(4,2)\simeq{\rm SU}(2,2).$$

 \section{ Spin$\mathbf{(V)}$ superalgebras}

 We now consider the embedding of ${\rm Spin}(V)$ in simple real superalgebras.
We require in general  that the odd generators are in a real
  spinor representation of
  ${\rm Spin}(V)$. In the cases $D_0=2,6$, $\rho_0=0,4$ we have to allow the two independent
irreducible  representations, $S^+$ and $S^-$ in the superalgebra, since  the
relevant morphism is
$$S^+\otimes S^-\longrightarrow \Lambda^2.$$
The algebra is then non chiral.

  We first consider minimal superalgebras, i.e. those with the
  minimal even subalgebra. From the classification of simple superalgebras \cite{nrs, ka} one
obtains the results listed in Table \ref{min}.

\begin{table}[ht]
\begin{center}
\begin{tabular} {|c|c|l|l|}
\hline
 $D_0$   & $\rho_0$& Spin($V$) algebra&Spin($V$) superalgebra\\\hline\hline
 1,7& 1,7& so($2^{(D-3)/2},2^{(D-3)/2}$)&  \\\hline
 1,7& 3,5 & so$^*$($2^{(D-1)/2}$)&osp$(2^{(D-1)/2})^*|2)$ \\\hline
 3,5& 1,7& sp($2^{(D-1)/2},\R$)&osp$(1|2^{(D-1)/2},\R)$\\\hline
 3,5& 3,5&
 usp($2^{(D-3)/2},2^{(D-3)/2}$)&osp$(2^*|2^{(D-3)/2},2^{(D-3)/2}$)
  \\\hline\hline
 0& 0&so($2^{(D-4)/2},2^{(D-4)/2}$) &  \\\hline
 0& 2,6&so($2^{(D-2)/2},\C)^\R$ &  \\\hline
 0& 4 & so$^*(2^{(D-2)/2}$)&osp$(2^{(D-2)/2})^*|2)$\\\hline
 2,6&  0&sl$(2^{(D-2)/2},\R)$&sl$(2^{(D-2)/2}|1)$ \\\hline
 2,6&2,6& su$(2^{(D-4)/2},2^{(D-4)/2})$&su$(2^{(D-4)/2},2^{(D-4)/2}|1)$\\\hline
2,6 & 4 &su$^*(2^{(D-2)/2}))$&su$(2^{(D-2)/2})^*|2)$ \\\hline
  4&0&sp($2^{(D-2)/2},\R$)&osp$(1|2^{(D-2)/2},\R)$ \\\hline
 4&2,6&sp($2^{(D-2)/2},\C)^\R$&osp$(1|2^{(D-2)/2},\C)$ \\\hline
 4&4&usp($2^{(D-4)/2},2^{(D-4)/2}$)&  osp$(2^*|2^{(D-4)/2},2^{(D-4)/2}$)\\\hline
\end{tabular}
\caption{Minimal Spin($V$) superalgebras.}\label{min}
\end{center}
\end{table}

We note that the even part of the minimal superalgebra contains the Spin($V$) algebra
 obtained in Section \ref{realforms} as a simple factor. For all quaternionic cases,
 $\rho_0=3,4,5$, a second simple factor SU(2) or SO$^*$(2) is present. For the linear cases there is
 an additional
non simple factor, SO(1,1) or U(1), as discussed in Section \ref{realforms}.

For $D=7$ and $\rho=3$ there is actually a smaller
 superalgebra, the exceptional  superalgebra  $f(4)$ with
 bosonic part  spin(5,2)$\times$su(2). The superalgebra
appearing in  Table \ref{min} belongs to the classical series and
its  even part is  so$^*(8)\times$su(2), being so$^*(8)$ the
Spin$(5,2)$-algebra.

Since we are considering minimal simple superalgebras, there are
some terms in the anticommutator that in principle are allowed
morphisms but that do not appear. One can see that these are
\begin{alignat*}{5}
D_0&=2,6,&\;\;\;\;\rho_0&=0,2,4,6,\;\;\;\;& &S^\pm \otimes S^\pm
\longrightarrow \sum_k\Lambda^{2k+1}, \\
D_0&=4,&\;\;\;\rho_0&=0,2,6,\;\;\;&&S^+\otimes S^- \longrightarrow
\sum_k\Lambda^{2k+1} ,
\\D_0&=1,7,&\;\;\;\rho_0&=3,5,\;\;\;&&S\otimes S\longrightarrow \sum
\limits_{k \neq 0}\Lambda^{4k},\\ D_0&=0, &\;
\;\;\rho_0&=4,\;\;\;& &S^+\otimes S^+\longrightarrow \sum
\limits_{k \neq 0}\Lambda^{4k}.
\end{alignat*}

Keeping the same number of odd generators, the  maximal simple  superalgebra
 containing ${\rm Spin}(V)$ is  an
 orthosymplectic algebra with ${\rm Spin}(V)\subset {\rm Sp}(2n,\R)$, being $2n$ the real
dimension of $S$. The various cases  are displayed in the Table
\ref{max}. We note that the minimal superalgebra is not a
subalgebra of the maximal one, although it is so for the bosonic
parts.

\begin{table}[ht]
\begin{center}
\begin{tabular} {|c|c|l|}
\hline
 $D_0$   & $\rho_0$& Orthosymplectic\\\hline\hline
3,5,& 1,7& osp$(1|2^{(D-1)/2},\R)$ \\\hline
 1,7& 3,5  & osp$(1|2^{(D+1)/2},\R)$ \\\hline
  0&4 &osp$(1|2^{D/2},\R)$  \\\hline 4&0& osp$(1|2^{(D-2)/2},\R)$
\\\hline
 4&2,6  &osp$(1|2^{D/2},\R)$ \\\hline
  2,6&  0&osp$(1|2^{D/2},\R)$ \\\hline
  2,6 & 4 & osp$(1|2^{(D+2)/2},\R)$\\\hline
 2,6&2,6&osp$(1|2^{D/2},\R)$ \\\hline
\end{tabular}
\caption{Maximal Spin($V$) superalgebras}\label{max}
\end{center}
\end{table}

Tables \ref{min} and \ref {max} show that there are 12 (mod(8) in $D$ and $\rho$)
 superalgebras for $D$ even
and 8 mod(8) superalgebras for $D$ odd, in correspondence with  Table \ref{con}.

\section{Summary}

In this paper we have considered superalgebras containing space-
time supersymmetry in arbitrary dimensions and with arbitrary
signature. In particular  super conformal algebras in $D$
dimensional space-time give, by Inon\"u- Wigner contraction, super
translation algebras with central charges in $D+1$ dimensions.
They also contain $D$ dimensional super Poincar\'e algebras as
subalgebras. The maximal central extension of the Poincar\'e
superalgebra  can be obtained by contraction of the ${\rm
osp}(1|2n,\R)$ superalgebra where n is related to the space
dimensions according to \ref{max}.

 In  Table \ref{lor} we report  these superalgebras for a
physical space-time of signature $(D-1,1)$, $D=3,\dots 12$. The first column
 (Lorentz) is the
supersymmetric extension the orthogonal algebra so$(D-1,1)$. The second column
 (Conformal) is the supersymmetric extension of the conformal algebra
  in dimension $D$, so$(D,2)$. The third column (Orthosymplectic) is
 the superalgebra that by contraction gives the maximal central
extension of the super translation algebra in dimension $D$. Note
that the same algebras appear in $D=3, 11$ and  in $D=4, 12$,
owing to the mod(8) periodicity.

\begin{table}[ht]
\begin{center}
\begin{tabular} {|c|c|c|c|}
\hline
 $D$   &Lorentz& Conformal &Orthosymplectic\\\hline\hline
3& osp$(1|2,\R)$ & osp$(1|4,\R)$ & osp$(1|2 ,\R)$ \\\hline
 4& osp$(1|2,\C)$  &su$(2,2|1)$ & osp$(1|4,\R)$ \\\hline
  5&  &osp$(8^*|2)$ & osp$(1|8,\R)$  \\\hline
6&su$(4^*|2)$&osp$(8^*|2)$ &osp$(1|16,\R)$\\\hline
 7&osp$(8^*|2)$  &osp$(16^*|2)$&osp$(1|16,\R)$  \\\hline
  8&  &su$(8,8|1)$ &osp$(1|32,\R)$ \\\hline
  9 &  &osp$(1|32,\R)$&osp$(1|32,\R)$ \\\hline
 10&sl$(16|1)$&osp$(1|32,\R)$ &osp$(1|32,\R)$ \\\hline
 11&osp$(1|32,\R)$&osp$(1|64,\R)$ &osp$(1|32,\R)$ \\\hline
 12&osp$(1|32,\C)$ &su$(32,32|1)$&osp$(1|64,\R)$ \\\hline
\end{tabular}
\caption{Supersymmetric
extensions of space-time groups.}\label{lor}
\end{center}
\end{table}

Note that the Poincar\'e supersymmetries obtained by contractions of the
 orthosymplectic algebras in Table \ref{lor} are non chiral for $D=6,10$ and
are $N=2$ for $D=8,9$. We can compare Table \ref{lor} with Table 7
of reference \cite{vv} \,\ dealing with $D=10,11,12$. We find
general agreement although in reference \cite{vv} the real forms
of the supergroups were not worked out. Furthermore in the case of
Lorentz superalgebra in $D=12$ our analysis shows that the result
is ${\rm osp}(1|32,\C)$. Note that in $D=4$ we get both, the
Wess-Zumino $N=1$ super conformal algebra \cite{wz} and by
contraction of ${\rm osp}(1|4, \R)$, the Poincar\'e superalgebra
with the domain wall central charge \cite{ds}.

 It is worthwhile to mention that from our
tables we can retrieve the super conformal algebras that do not
violate the Coleman-Mandula \cite{cm} theorem and its
supersymmetric version, the Haag-Lopusza\'nsky-Sohnius theorem
\cite{hls}. These
 state that the even
part of the superalgebra should be given by the sum of the
space-time symmetry algebra and an internal symmetry algebra. It
is immediately  seen that this happens only for $D=3,4, 6$.
Indeed, this occurs because of the following isomorphisms:
\begin{equation*}
  {\rm SO}(3,2) \simeq {\rm Sp}(4,\R);\,\  {\rm SO}(4,2) \simeq {\rm SU}(2,2);\,\ {\rm SO}(6,2) \simeq {\rm
  SO}^*(8).
\end{equation*}
The $D=5$ case is also allowed if we replace the
${\rm osp}(8^*|2)$ superalgebra with the exceptional superalgebra
${\rm f}(4)$. The first departure occurs at $D=7$,
where the conformal group ${\rm SO}(7,2)$ must be embedded in
${\rm SO^*}(16)$ to find a supersymmetric extension.

\bigskip

\noindent{\bf \Large Appendix A}

\bigskip

Some embeddings of real forms of non compact groups which have
been used through the text are given below.

\begin{eqnarray*}
 {\rm SO}(p+q,\C)_\R& \supset& {\rm SO}(p,q)\\
{\rm SO}(2n,\C)_\R& \supset& {\rm SO^*}(2n)\\ {\rm SO}(n,n)&
\supset& {\rm SO}(n,\C)_\R\\
   {\rm SO}(n,n)& \supset& {\rm SL}(n,\R) \times
   {\rm SO}(1,1)={\rm GL}(n,\R)\\
     {\rm SO}(4n,4n)& \supset& {\rm SU}(2) \times {\rm Usp}(2n,2n)\\
    {\rm SO^*}(2p+2q)& \supset& {\rm SU}(p,q)\times
   {\rm U}(1)\\{\rm SO^*}(2n)& \supset& {\rm SO}(n,\C)_\R\\
  {\rm SO^*}(4n)& \supset& {\rm SU^*}(2n)\times
   {\rm SO}(1,1)\\
{\rm Sp}(2p+2q,\C)_\R& \supset& {\rm Usp}(2p,2q)\\ {\rm
Sp}(2p+2q,\R)& \supset& {\rm U}(p,q)\\
 {\rm Sp}(2n,\R) & \supset & {\rm GL}(n,\R)\\
    {\rm Sp}(4n,\R)& \supset& {\rm Sp}(2n,\C)_\R\\
 {\rm Sp}(4n,\R)& \supset& {\rm SU}(2) \times {\rm SO^*}(2n) \\
{\rm Usp}(2n,2n)& \supset& {\rm SU^*}(2n)\times
   {\rm SO}(1,1)\\
\end{eqnarray*}

\bigskip

\noindent{\bf \Large Appendix B}

\bigskip

We give explicitly the decomposition of the tensor product
representation $S\otimes S$.
 The Clifford algebra has a $\Z_2$ grading,
$$\Ca(s,t)=\Ca^+(s,t)\oplus \Ca^-(s,t),$$
where
$$\Ca^+(s,t)=\sum_{k}\Lambda^{2k},\qquad \Ca^-(s,t)=\sum_{k}\Lambda^{2k+1}.$$

\paragraph{$\mathbf{D}$ odd.} $\Ca^\pm$ carry isomorphic
 representations of ${\rm Spin}(s,t)$, since $\Lambda^k\approx \Lambda^{D-k}$.
 We can consider only $\Ca^+$. We have then
$$\Ca^+=\A_0+\A_2, \qquad \A_0=\sum_{k}\Lambda^{4k}, \;\; \A_2= \sum_{k}\Lambda^{4k+2}.$$
Form Table \ref{morphisms} it follows that the morphisms in $\A_0$ are symmetric
 for $D_0=1,7$ and antisymmetric for $D_0=3,5$. $\A_2$ is symmetric for $D_0=3,5$ and
antisymmetric for $D_0=1,7$.

\paragraph{$\mathbf{D}$ even.} $\Ca^+$ is not isomorphic to $\Ca^-$.
\begin{eqnarray*}
\Ca^+=\A_0+\A_2, \qquad \A_0=\sum_{k}\Lambda^{4k}, \;\; \A_2= \sum_{k}\Lambda^{4k+2},\\
\Ca^-=\A_1+\A_3, \qquad \A_1=\sum_{k}\Lambda^{4k+1}, \;\; \A_3= \sum_{k}\Lambda^{4k+3}.
\end{eqnarray*}
The symmetry properties are given in Table \ref{pitab}.
\begin{table}[t]
\begin{center}
\begin{tabular} {|c|c|c|c|c|}
\hline
    &$D_0=0$&$D_0=2$ &$D_0=4$&$D_0=6$\\\hline\hline
$\A_0$& $+$ & $\pm$&$-$&$\mp$ \\\hline
$\A_1$& $\pm$ & $+$&$\mp$&$-$ \\\hline
$\A_2$&  $-$ & $\mp$&$+$&$\pm$ \\\hline
$\A_3$& $\mp$ & $-$&$\pm$&$+$ \\\hline
\end{tabular}
\caption{Symmetries of gamma matrices for $D$ even \label{pitab}}
\end{center}
\end{table}

$+$ means symmetric and
 $-$ antisymmetric. $\pm$ and $\mp$ are symmetric or antisymmetric depending on
the choice of the charge conjugation matrix (see Section
\ref{raja}).

The morphisms are
\begin{eqnarray*}
S^\pm\otimes S^\pm\longrightarrow \Ca^+, \quad D=0,4\\
S^\pm\otimes S^\pm\longrightarrow \Ca^-, \quad D=2,6\\
S^\pm\otimes S^\mp\longrightarrow \Ca^+, \quad D=2,6\\
S^\pm\otimes S^\mp\longrightarrow \Ca^-, \quad D=0,4
\end{eqnarray*}

\bigskip

The Spin($V$)-algebra is the module $\A_2$. The compact generators for the case of
 Minkowskian signature $(D-1,1)$, are given by the space like
components of the even generators,  $Z_{[i_1\dots i_k]}$,
 $i_j=1,\dots D-1$. The maximal compact subgroups are
\begin{eqnarray*}
&D=3\qquad & {\rm U}(1)\\ &D=4\qquad & {\rm SU}(2)\\ &D=5\qquad &
{\rm SU}(2)\times {\rm SU}(2)\\ &D=6\qquad & {\rm USp}(4)\\
&D=7\qquad & {\rm U}(4)\\ &D=8\qquad & {\rm SO}(8)\\ &D=9\qquad &
{\rm SO}(8)\times {\rm SO}(8)\\ &D=10\qquad & {\rm SO}(16)\\
&D=11\qquad & {\rm U}(16)\\ &D=12\qquad & {\rm USp}(32).
\end{eqnarray*}

This is in agreement with Table \ref{min}.

\bigskip

\noindent {\bf \Large Acknowledgements.}

\bigskip

 V. S. V. would like to
express his gratitude to the Dipartimento di Fisica, Universit\`a
di Genova and the INFN, Sezione di Genova, for their hospitality
during his stay in September 2000, when most of the work for this
paper was done. S. F. would like to thank A. Van Proeyen for
useful discussions and comments.  R. D. and M. A. Ll. want to
thank the Theory Division of CERN for its kind hospitality during
the elaboration of this work. The work of R. D. has been supported
in part by the European Commission RTN network HPRN-CT-2000-00131
(Politecnico di Torino). The work of S. F. has been supported in
part by the European Commission RTN network HPRN-CT-2000-00131,
(Laboratori Nazionali di Frascati, INFN) and by the D.O.E. grant
DE-FG03-91ER40662, Task C.

\end{document}